\documentclass[twocolumn,showpacs,preprintnumbers,amsmath,amssymb]{revtex4}

\def\dt{{\tilde \delta}}

\def\kt{\tilde k}

\usepackage{graphicx}
\usepackage{dcolumn}
\usepackage{bm}

\begin{document}

\title{The Fermi surface reconstruction in stripe phases of
cuprates}

\author{M.\,Ya.\,Ovchinnikova}
\affiliation{N.N.Semenov Institute of Chemical Physics RAS, 117334
Moscow, Russia}

\begin{abstract}
Mean-field study of the stripe structures is conducted for a
hole-doped Hubbard model. For bond-directed stripes, the Fermi
surface consists of segments of an open surface and the boundaries
of the hole pockets which appear in the diagonal region of momenta
under certain conditions. Segments of the first type are due to
one-dimensional bands of states localized on the domain walls. The
relation of bands to the doping and temperature dependences of the
Hall constant is discussed. In connection with the observation of
quantum magnetic oscillations, a systematic search for the electron
pockets has been carried out. It is shown that the formation of such
pockets in bilayer models is quite possible.
\end{abstract}


\maketitle

Numerous properties of  underdoped (UD) cuprates are connected with
the formation of stripes - the periodic spin and charge structures
specified from the positions of magnetic peaks in neutron scattering
\cite{1}. New impetus to a discussion of the normal state properties
has been given by the measurements of longitudinal and Hall
conductivities of cuprates in strong magnetic fields suppressing the
superconductivity. A change of sign of the Hall coefficient $R_H$ as
$T\to 0$ in the $Y$- and $La$- based cuprates \cite{1a,1b} and
magnetic quantum oscillations of $R_H$, magnetization and
conductivity have been revealed in $YBa_2Cu_3O_{6+y}$ (Y123) and
$YBa_2Cu_4O_{8}$ (Y124) compounds \cite{1a,2,3}.

The observations testify to a reconstruction of the Fermi surface
(FS). The temperature of its onset is associated with the start of
deviations of resistivities and the Hall coefficient $R_H(T)$ from a
linear $T$-dependence typical for large $T$. There are two doping
regions in UD cuprates with different signs of the deviations. First
doping region near an optimal doping reveals itself in a sharp
increase of positive $R_H(T)$ for $T<T^*$ in $La_{1-x}Sr_xCu_4$
(LSCO) \cite{3a}. Such behavior may be due to the transition from
paramagnetic normal state with a large FS  to the state with some
charge-spin structure exemplified by dividing the large FS into arcs
of the hole pockets. In the region of small doping, the $R_H(T)$
falls down for $T<T^*$ and becomes negative as $T\to 0$. This might
be connected with changes in the charge and spin stripes, which are
most evident near the temperature $T_s$ of the structural $LTO\to
LTT$ transition between low-temperature orthorhombic (LTO) and
low-temperature tetragonal (LTT) phases. To explain the negative
sign and quantum oscillations of $R_H$ as $T\to 0$, it was supposed
that an opening of the electron (e-) pockets in FS occurs
\cite{2,3}. Then a question may be posed: can the stripe phases give
rise to such a sort of FS segments? In principle, work \cite{4}
confirms such a possibility. But it leans upon a very rough model in
which the stripe impact on electronic states is described by an
outer periodic potential.

As distinct from two Hubbard bands of a homogeneous
antiferromagnetic state, a periodic structure with $n_c$ sites in a
unit cell is characterized by $n_c$ bands. Their sections at the
Fermi level provide the FS segments of different types. To study
them, one must be convinced of the stripe formation and determine
the self-consistent periodic potential.

In the present work, the mean field (MF) approximation is used to
study the bands and the FSs of periodic stripe structures in the
t-t'-t"-U Hubbard models. Earlier MF studies \cite{5,5a,5b} showed
that the stripes provide an explaination for the observed
fragmentation of FSs and dichotomy of the nodal and antinodal
quasiparticles. Here we extend the calculations to a more wide set
of structures with the systematic search for electronic pockets
among the FS segments.

The zero band of the model is taken in the form
$\epsilon(k)=-2t(c_x+c_y)+4t'c_xc_y-4t''(c_x^2+c_y^2-1)$, where
~~$c_{x(y)}=\cos{[k_{x(y)}]}$. The on-site repulsion and positive
values of intersite hopping at the distances $a,~\sqrt{2}a,~ 2a$
($a$ is the lattice constant) are equal to
\begin{equation}
U/t=4, ~~t'/t=0.3,~~t''/t'=0.5
\label{1}
\end{equation}
Our calculations will cover the doping range with $x>0.05$ and the
structures $S_l$ or $B_l$ with the domain walls (DWs) directed along
the $y$-bonds and centered at the sites or the bonds, respectively.
The stripe period, i.e., the distance between DWs $l$ (in units of
$a$) is varied in the range $l=7\div 4$. This range corresponds to
the observed linear doping dependence of the incommensurability
parameter $\delta={\frac{1} {2l}}=x$ for $x\le {\frac{1} {8}}$ and
its saturation value $\delta={\frac{1}{8}}$ for $x>{\frac{1}{8}}$.
The direct and reciprocal lattice vectors of the structure are
$E_{1,2}=(la, \pm a)$,
~$B_{1,2}={\frac{\pi}{a}}({\frac{1}{l}},\pm1)$ or $E_1=(la,0)$,
$E_2=(0,2a)$ ,~$B_1=({\frac{2\pi}{la}}, 0)$,
~$B_2=(0,{\frac{\pi}{a}})$ at even $l$ or odd $l$, respectively. A
number of sites in a unit cell equals $n_c=2l$.

The order parameters of a periodic MF solution are represented by
the charge and spin densities at the  sites of a unit cell. The MF
procedure was described in Refs. \cite{5,5b}. One-particle states
\begin{equation}
\chi_{\kt\nu}^\dagger=\sum_{m,\sigma}c^{\dagger}_{{k_m},\sigma}W_{m\sigma,\nu}(\kt)
\label{2}
\end{equation}
are the eigenstates of a linearized Hamiltonian. They determine the
spin-degenerate bands $E_\nu(\kt)$ ($\nu=1,\ldots n_c$) of the
periodic structure. Here, $k_m=\kt+Bm$, ~$Bm=B_1 m_1+B_2 m_2$,
~~$\kt$ runs over momenta in a reduced Brillouin zone (BZ) of the
structure, and $m=(m_1,m_2)$ numerates all independent Umklapp
vectors. The $E_{\nu}(k)$ bands, the $W$ matrix in formula (2) and
the Fermi function $f(\omega)$ determine the spectral density
$A(k,\omega)$ and the photoemission intensity $I(k,\omega)$:
\begin{equation}
\begin{array}{ll}
&I(k,\omega) \sim A(k,\omega)f(\omega); \\
&A(k,\omega) =\sum_{m,\nu,\sigma} |W_{m\sigma,\nu}|^2 \dt(E_\nu(k_m)-\mu-\omega)\\
\end{array}
\label{3}
\end{equation}
A density map of function (3) at $\omega=0$ visualizes the main and
shadow segments of the FS. A density map of the local (in k-space)
density of states, viz.
\begin{equation}
D(k,\omega)=\sum_\nu\dt(E_{k,\nu}-\mu-\omega),
\label{4}
\end{equation}
allows us to overview full sections of bands at the level
$E=\mu+\omega$ regardless of their weights in photoemission. Here,
$\dt$  in Eqs (3),(4) is the delta-function with some finite width
which was taken $\sim (0.02\div 0.04)t$ by order of magnitude.
\begin{figure}
\includegraphics[width=.47\textwidth]{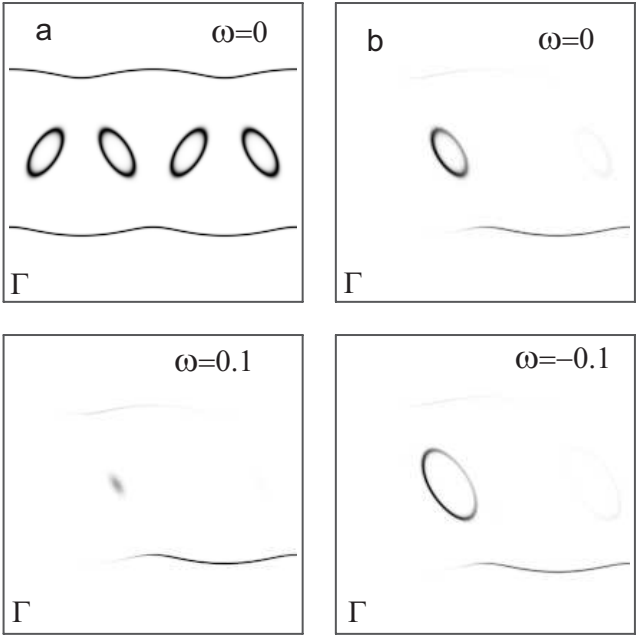}
\caption{Fig.1. Fermi surfaces and their 'visible' segments for the
$S_4$ structure with $l=4$ at doping $x=0.15$ on the density maps of
(a) the local density of states $D(k,\omega=0)$, and (b) the
spectral function $A(k,\omega=0)$ in the first quadrant of a
Brilluoin zone. Low panels: the density maps of $A(k,\omega)$ at
$\omega=\pm 0.1$ (in units of $t$).}
\end{figure}

Figure 1 presents typical FSs and their 'visible' segments for the
$S_4$ structure at doping $x=0.15$. It has been checked that the
open FSs give rise to quasi-one-dimensional (1D ) bands of states
localized on the domain walls. Closed segments of FSs represent the
boundaries of the hole pockets. A decrease in doping leads to
narrowing the h-pocket. It is clearly seen from the maps of the
spectral function at $\omega\neq 0$ in Fig.1. For the same $S_4$
structures, the hole pockets disappear for $x<x_{arc}=0.135$ and
only 1D FS segments retain. Similar behavior is displayed by the
$B_4$ structure.

The 1D bands with a full suppress of photoemission from the nodal
region have been observed in $La_{1.28}Nd_{0.6}Sr_{0.12}CuO_4$
compound \cite{6} in which the static stripes have been revealed.
The temperature and doping dependences of the Hall constant have
been explained \cite{7} by a crossover from the 2D to 1D type
electronic structure which occurs with disappearing the h-pockets.

Figure 2 depicts the energies $[E_{\nu}(k_x,k_y)-\mu]$ of bands
$\nu=3,4,5$ closest to the Fermi level $\mu$ for the $S_4$ model at
doping $x=1/8$. Numbering of bands is made in increasing order of
their energies. Band $\nu=3$ is close to touch the Fermi level from
below and, consequently, it is responsible for the hole pockets at
higher doping. Band $\nu=l=4$ is the 1D band with an open FS. And
the band $\nu=5$ approaches the Fermi level from above at the points
$k=(0,\pi)+B_i m_i$. Decrease of doping at the fixed stripe period
$l$ would led to the appearance of electronic pockets. But in
reality it is accompanied by a change in the stripe period $l$
according to ${\bar l}(x)\sim 1/2x$ at low $T$ and $x<1/8$.
\begin{figure}
\includegraphics[width=.47\textwidth]{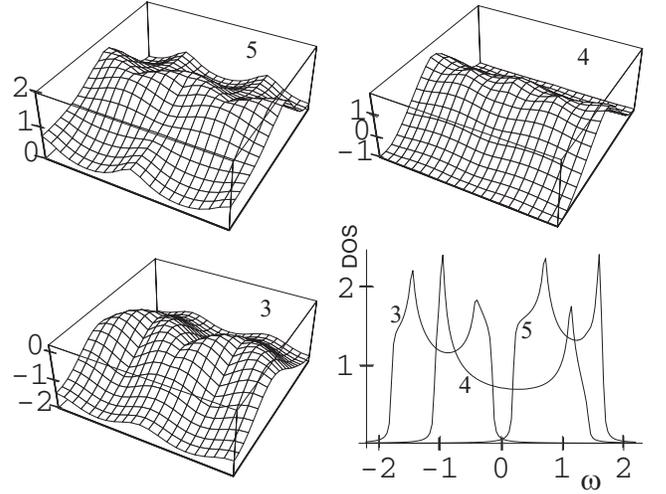}
\caption{Figure 2. Energies $[E_{\nu}(k)-\mu]$ for bands $\nu=3,4,5$
of the $S_4$ structure at doping $x=1/8$ in the first quadrant of
BZ. Right bottom: the $DOS_{\nu}(\omega)$ contributions from the
same bands into the density of states (per one unit cell). The band
numbers $\nu$ mark the graphics and curves. The energies and
$\omega$ are given in units of $t$.}
\end{figure}

Characteristic cyclotron masses $m_1$ and $m_2$ of the h- or
e-carriers in the case of opening the corresponding h- or e-pockets
or during their activation depend only slightly on doping and are of
order $m_1\sim 0.41m_e$ and $m_2\sim 0.92m_e$ ($m_e$ is the electron
mass).

Notice that the self-organized stripes lead to formation of some gap
in the density of states $DOS(\omega)$ of the system at the Fermi
level. It is not a full gap [$DOS(\omega=0)\ne 0$]. But such a gap
is absent in the homogeneous antiferromagnetic state of model (1):
contributions to the $DOS(\omega)$ from upper and lower Hubbard
bands are overlapping. Situation is similar to that in
semiconductors where the self-organized charge distribution among
impurities provides a gap in $DOS$ at the Fermi level.

Thus, the $S_l,~B_l$ structures with the period $l=5,6,7$ correspond
to doping range $0.05\leq x\leq 0.125$ at small $T$. At doping
$x_l=1/2l$ there are only 1D segments of the FS. The hole pockets
and nodal Fermi arcs are absent, though the system is close to their
opening. Variations in the temperature for the fixed structure
change only slightly the bands and FSs of the MF solutions.
Therefore, the observed temperature dependences of the Hall constant
and the length of the small Fermi arc cannot be explained in the
context of a fixed structure. They may be understood if one infers a
change in structure with increasing $T$, for example, a growth of
the average stripe period with $T$. The growth of ${\bar l}(T)$
decreases a density of domain walls and a concentration of holes
localized on them. At fixed doping, this is accompanied by the
opening of the h-pockets, increase of the Fermi-arc length, and
growth of the positive (hole) contribution to the Hall constant.
Hypothesis for the $T$-dependent ${\bar l}$ is in accordance with
the observed temperature dependence of the incommensurability
parameter, which shows up more explicitly in the region near a
transition from tetragonal to orthorhombic ($LLT\to LTO$)  phases
in $La_{2-x}Ba_xCuO_4$  
\cite{8} at doping $x=1/8$.

It should be noted that the quasi-1D FS segments in LSCO (unlike
NLSCO) are not observed in the angle-resolved photoemission spectra
(ARPES) \cite{9}. This may be connected with the irregular stripes,
inhomogeneity and multi-structure nature of LSCO.

The results obtained for periodic structures may be usefull in
interpreting the two-component character of the effective
concentration $n_H(x,T)$ of the Hall carriers \cite{10}. Its doping
and temperature dependences have the following form \cite{10,11,12}
\begin{equation}
n_H=n_0(x)+n_1(x)\exp{[-E(x)/kT]}.
 \label{5}
\end{equation}
Concentration $n_0(x)=x$ of the first type carriers which survive as
$T\to 0$ coincides with the concentration $n_h=1/2l=x$ of holes
localized on the domain walls if the hole pockets have not yet
opened (for $x<x_{arc}$). The second type carriers in formula (5)
may originate from the activation of holes in the nodal region
before the opening of the hole pockets at small $T$ and $x$. At
large $T$, after the opening of h-pockets, these carriers might
originate from a large Fermi arc including the antinodal regions of
the van-Hove singularity which resides below the Fermi level in UD
cuprates.

\begin{figure}
\includegraphics[width=.47\textwidth]{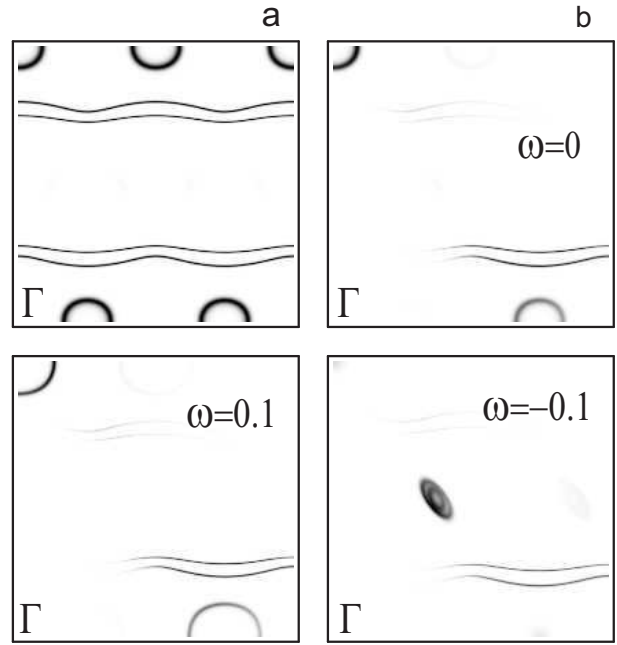}
\caption{Figure 3. Fermi surfaces and their 'visible' segments on
the maps of the local density of states $D(k,\omega=0)$ and the
spectral density $A_{\Sigma}(k,\omega=0)$ for bilayer model with
parameters (1), $t_z=0.35t$ and the stripe structure $S_4$ in each
layer. Lower panels: the maps of $A_{\Sigma}(k,\omega)$ at
$\omega=\pm 0.1t$ demonstrate evolution of e- and h-pockets. All
maps are given in I quadrant of BZ at doping 0.125.}
\end{figure}

The  monolayer models studied above do not display the electron
pockets. A decrease in the most crucial parameter $t'/t$ down to the
value $t'/t=0.2$ strongly increases the activation energy for
quasiparticles from the potential e-pockets: the gap between a
quasi-1D band $\nu=l$ and next upper band $\nu=l+1$ for the
$S_l,B_l$ structures sharply increases with a decrease in $t'/t$.
The 1D bands may provide the negative contribution to the Hall
conductivity $\sigma_H$. This was confirmed by calculations of the
1D band contribution in a first approximation with respect to the
overlapping between the states localized on the neighboring domain
walls. The estimates have been made in the context of hopping
mechanism of conductivity, which is  suitable at small $T$. But the
1D bands with open FS cannot explain the magnetic quantum
oscillations.

Taking into account the above reasoning we remind that quantum
oscillations have been observed in the bilayer cuprates, viz. in the
ortho-II Y123 and Y124 \cite{2,3}. Therefore, we have calculated the
FSs and bands of stripes in bilayer models. Interlayer interaction
is then described by a splitting
$\Delta=t_z(\cos{k_x}-\cos{k_y})^2/2$ of the zero bands in the
standard form \cite{12}. Figure 3 presents the FSs for bilayer model
with the parameters entering Eqn (1) and $t_z/t=0.35$, and
synchronized antiphase stripe structures in both layers
($S_{zn1}=-S_{zn2}$ in the layers 1,2). A map of the summary local
density of states $D_{\Sigma}(k,\omega)$ and the spectral function
$A_{\Sigma}(k,\omega)=A_{+}(k,\omega)+ A_{-}(k,\omega)$ from the
bonding and antibonding bands (BB and AB, respectively) are given at
doping $x=0.125$. For the stripes directed along $y$, the e-pockets
are seen at the points $(0,\pi)+ B_i m_i$ in addition to the split
open FS segments of 1D bands. The boundaries of the e-pockets around
$(0,\pm\pi)$ display the maximum photoemission intensity, and they
correspond to BB bands. The electronic nature of these pockets
(unlike the nodal hole pockets) is proved by their evolution with
$\omega$ on the map $A_{\Sigma}(k,\omega)$ (fig. 3).

Although the zero band bilayer splitting goes to zero in the nodal
lines, despite that the bilayer splitting of the 1D bands is only
slightly dependent on the direction of the vector $[k-(\pi,\pi)]$
for $k$ moving along FSs of these bands. The area of electronic
pocket represents about $1.1\%$ of the magnetic Brillouin zone area.
The order of magnitude of this value does not contradict to the
estimate of $2.3\%$ obtained from the observed frequency of quantum
magnetic oscillations.

{\it In summary,} the MF calculations of the stripe structures in UD
Hubbard model reveal two types of the Fermi surface segments:
segments of open FSs from the 1D bands, and closed boundaries around
the hole pockets in the nodal region. First type carriers with the
concentration linear in doping refer to states localized on the
domain walls and provide the quasi-1D charge transport. For the
observed relation $x=1/2{\bar l}(x)\leq 1/8$ between doping and the
stripe period only the first type carriers survive as $T\to 0$ with
a small contribution from activated  hole carriers. Increase in the
activated hole concentration with increasing $T$ explains the rise
of $R_H$ from negative to positive values if one infers the growth
of the stripe period with $T$. Negative $R_H$ as $T\to 0$ might be
caused by hoping conductivity of the 1D carriers. But quantum
oscillations can only be explained by the appearance of e-pockets.
In monolayer models, the search for them fails. But the bilayer
striped models exhibit the opening of electronic pockets in
antinodal regions for bonding bands. Just these e-pockets might be
responsible for magnetic quantum oscillations in YBCO.

The author is thankful to V.Ya.Krivnov for taking interest in the
problem and useful discussions.

\end{document}